\documentclass[sigconf]{acmart}
\usepackage{natbib}
\usepackage{multirow}
\usepackage{booktabs}      
\usepackage{multirow}      
\usepackage{enumitem}      
\usepackage{tabularx}
\usepackage{subcaption} 
\usepackage{tikz}
\usetikzlibrary{positioning,arrows.meta,shapes}
\usepackage{graphics}
\usepackage{pifont}
\newcommand{\xmark}{\ding{55}} 

\usepackage{xcolor}
\definecolor{blue}{rgb}{0,0,0}
\AtBeginDocument{%
  }

\setcopyright{acmlicensed}
\copyrightyear{2018}
\acmYear{2018}
\acmDOI{XXXXXXX.XXXXXXX}
\acmConference[Conference acronym 'XX]{Make sure to enter the correct
  conference title from your rights confirmation email}{June 03--05,
  2018}{Woodstock, NY}
\acmISBN{978-1-4503-XXXX-X/2018/06}

\begin{document}

\title{Shades of Uncertainty: How AI Uncertainty Visualizations Affect Trust in Alzheimer’s Predictions}

\author{Jonatan Reyes}
\email{jonatan.reyes01@utrgv.edu}
\orcid{0000-0002-8377-5244}
\authornotemark[1]
\affiliation{%
  \institution{The University of Texas Rio Grande Valley}
  \city{Edinburg}
  \state{Texas}
  \country{USA}
}

\author{Mina Massoumi}
\affiliation{%
 \institution{Concordia University}
 \city{Montr\'eal}
 \state{Qu\'ebec}
 \country{Canada}}
 
\author{Anil Ufuk Batmaz}
\affiliation{%
 \institution{Concordia University}
 \city{Montr\'eal}
 \state{Qu\'ebec}
 \country{Canada}}

\author{Marta Kersten-Oertel}
\affiliation{%
 \institution{Concordia University}
 \city{Montr\'eal}
 \state{Qu\'ebec}
 \country{Canada}
}

\renewcommand{\shortauthors}{Trovato et al.}

\begin{abstract}

\textcolor{blue}{Artificial intelligence (AI) is increasingly used to support prognosis in Alzheimer’s disease (AD), but adoption remains limited due to a lack of transparency and interpretability, particularly for long-term predictions where uncertainty is intrinsic and outcomes may not be known for years. We position \emph{uncertainty visualization} as an explainable AI (XAI) technique and examine how it shapes trust, confidence, and reliance when users interpret AI-generated forecasts of future cognitive decline transitions. We conducted two studies, one with general participants (N=37) and one with experts in neuroimaging and neurology (N=10), to compare binary (present/absent) and continuous (saturation) uncertainty encodings. Continuous encodings improved perceived reliability and helped users recognize model limitations, while binary encodings increased momentary confidence, revealing expertise-dependent trade-offs in interpreting future predictions under high uncertainty. These findings surface key challenges in designing uncertainty representations for prognostic AI and culminate in a set of empirically grounded guidelines for creating trustworthy, user-appropriate clinical decision support tools.}
\end{abstract}



\maketitle

\section{Introduction}

Alzheimer’s Disease (AD) affects over 50 million people worldwide, a number projected to triple by 2050~\cite{winblad2016defeating,prince2015world}. Because there is no cure, early detection and accurate prognosis are essential for timely intervention, patient counseling, and clinical trial recruitment~\cite{better2023alzheimer}. \textcolor{blue}{A central clinical challenge is understanding how different biomarkers interact with one another and relate to disease development so that  \textit{disease progression trajectories} can be accurately discriminated and predicted from prodromal stages~\cite{orru2012using,eslami2023unique,tan2020longitudinal}}. Predicting these disease progression trajectories, such as the transition from cognitively normal (CN) to mild cognitive impairment (MCI), or from MCI to AD, is difficult because they unfold slowly, heterogeneously, and often years before overt symptoms~\cite{eskildsen2013prediction}. Yet these transitions carry major consequences for treatment planning and risk communication.


 \textcolor{blue}{Artificial Intelligence (AI) models have shown promise for modeling AD progression by integrating multimodal data including magnetic resonance imaging (MRI) and positron emission tomography (PET) imaging, cerebrospinal fluid (CSF) biomarkers, genetic risk factors, and cognitive assessments~\cite{tanveer2020machine,jo2019deep,zhao2023application}.} Such models can detect subtle neurodegeneration patterns and generate long-term prognostic forecasts that may support clinical decision-making. However, despite technical progress, clinical adoption remains limited, in large part due to the risk of \textcolor{blue}{misinterpreting AI outputs and their associated uncertainty}. When predictions for different levels of disease severity appear visually similar, clinicians may misjudge how confident the model actually is, leading to over- or under-reliance on its forecasts.

A major barrier for AI-based systems in AD is that clinical and imaging data are heterogeneous, irregularly sampled, and frequently incomplete due to missing visits, inconsistent follow-up, and variable imaging availability, producing substantial uncertainty in long-term predictions~\cite{petersen2010mild,schmidt2018missing}. Thus both aleatoric uncertainty (noise and variability in biomarkers and cognitive measures) and epistemic uncertainty (gaps in longitudinal records, imbalanced populations, and limited representation of atypical trajectories)~\cite{kendall2017uncertainties,reyes2020interpretability} are high. Clinicians and researchers typically receive categorical predictions or confidence scores from AI models without insight into \emph{why} the model is uncertain or how reliable a trajectory prediction might be~\cite{mittelstadt2019explaining}. This opacity complicates trust calibration, risking both over-reliance and under-reliance in high-stakes settings~\cite{dietvorst2015algorithm,bansal2021does}. {\textcolor{blue}{This creates a persistent gap between how models present uncertainty and how clinicians interpret it, making it difficult to assess appropriate trust in AI predictions.} 

\textcolor{blue}{Although AI research for AD increasingly emphasizes predictive performance, relatively few studies address the need for trustworthy, interpretable, and explainable outputs that can meaningfully support clinical adoption. Ideally, AI systems should help clinicians understand how domain-specific features extracted from heterogeneous clinical data contribute to suggested decisions or diagnoses. Recent systematic reviews of explainable AI (XAI) in AD research highlight that models often rely on attribution- and occlusion-based explanation techniques, as well as interpretable model architectures, to expose aspects of their decision logic~\cite{el2023trustworthy, viswan2024explainable}. However, most of these works focus on short-term or diagnostic tasks, while long-term prognosis, where uncertainty is both more substantial and more consequential, remains comparatively underexplored. Moreover, prior research provides limited insight into how different stakeholders interpret these explanations, including how clinicians and generalists make sense of uncertainty visualizations in high-stakes medical contexts~\cite{ghassemi2021false,al2025artificial,gonzalez2023scoping,kumar2025enhancing}.}

\textcolor{blue}{At the same time, research on uncertainty communication highlights substantial barriers to designing and implementing effective uncertainty visualizations~\cite{vazquez2020aggregation}, including constraints in methodological approaches and persistent challenges in end-user interpretability. Other studies note that visualization techniques remain underutilized in many existing AI systems despite their potential to support user understanding~\cite{eslami2023unique,tuauctan2021artificial,khojaste2022deep}. A recent survey further argues that current literature does not adequately address trustworthiness~\cite{el2023trustworthy} and calls for combining XAI methods with appropriate data representations to communicate model complexity, conflicts, and uncertainties more effectively. Building on these insights, we argue that uncertainty visualization, used as an XAI technique, offers an opportunity to make model confidence and limitations transparent to clinicians and other stakeholders, particularly in high-uncertainty prognostic tasks~\cite{doula2022visualization,cassenti2023representing}.}

\textcolor{blue}{This paper addresses these gaps by exploring how uncertainty visualizations shape trust, confidence, and decision-making in AI-supported AD progression prediction. AD prognosis represents a uniquely challenging setting for uncertainty communication. Predictions must extrapolate months or years into the future, giving clinicians no immediate feedback with which to recalibrate trust~\cite{salimzadeh2024dealing}. Diagnostic categories such as MCI remain debated, patient trajectories vary widely, and missing or irregular data are common. Errors can meaningfully affect care planning and trial eligibility. Collectively, these factors create a fragile decision-making environment in which communicating uncertainty is essential for calibrated reliance, motivating our investigation into how uncertainty visualizations affect trust and decision-making. Across two complementary user studies, we provide the following contributions: 
\begin{itemize}
    \item \textbf{XAI:} We position uncertainty visualization as an XAI technique.
    \item \textbf{XAI/HCI:} We present an exploratory analysis of how trust in AI is calibrated when users are provided with information about the model’s training and validation steps in advance.
    \item \textbf{AD:} We extend existing AD research by examining how generalist users and domain experts perceive AI-generated outputs with uncertainty information and how it shapes trust calibration, perceived reliability, and confidence when interpreting predicted CN/MCI/AD transitions.
    \item \textbf{HCI:} We provide design guidelines for uncertainty-aware clinical AI tools, grounded in quantitative patterns and qualitative insights from both studies.
\end{itemize}
Together, these contributions advance understanding of how uncertainty visualization functions as a form of explainability in high-stakes, long-term medical AI and offer practical guidance for designing trustworthy decision-support tools for AD prognosis.}



\section{Related Work}

Human–AI collaboration in healthcare involves complex interactions among system design, task characteristics, and user expertise. These factors influence whether clinicians adopt, defer to, or override algorithmic recommendations, with direct implications for patient safety~\cite{abujaber2022enabling,wang2021factors,bertao2021artificial,dietvorst2020people,park2022family}. In this section, we review three areas of prior work most relevant to our study: (1) AI-based clinical decision support, (2) trust and reliance in human–AI collaboration, and (3) uncertainty visualization in AI systems, with particular attention to AD prognosis.

\subsection{AI-Based Clinical Decision Support}

Clinical decision support systems (CDSS) integrate patient data into algorithmic recommendations to enhance diagnostic efficiency, safety, and workflow~\cite{kawamoto2005improving,sutton2020overview}. However, these benefits come with risks such as automation bias, where clinicians may over-rely on AI output—even when it is incorrect—particularly under time pressure or cognitive load~\cite{sutton2018much}. Empirical findings show that experts sometimes defer to erroneous AI suggestions in computational pathology~\cite{rosbach2024}, COVID-19 triage~\cite{wysocki2023assessing}, and clinical pharmacy~\cite{salim2023}, illustrating that training and domain expertise alone do not guarantee well-calibrated use of AI.

This literature highlights a central requirement for safety-critical AI: systems must support \textit{calibrated} trust rather than simply increasing trust~\cite{buccinca2021trust}. Proposed mechanisms include uncertainty communication~\cite{padilla2020uncertainty}, transparent explanations~\cite{ribeiro2016should}, interactive tools that convey system limitations~\cite{he2023users, liao2020questioning}, and decision-focused feedback loops~\cite{ahn2024impact}. Yet for prognostic tasks, where outcomes may not be observable for years, feedback is rarely available, making prospective uncertainty communication especially important.

\subsection{Trust, Reliance, and Task Characteristics in Human–AI Collaboration}

Trust reflects a user’s belief in an AI system’s competence, while reliance captures their observable use of system recommendations~\cite{lee2004trust,salimzadeh2024dealing}. These constructs often diverge: users may trust a system yet avoid relying on it, or rely on a system despite limited trust~\cite{dzindolet2003role}. Methods for measuring trust and reliance include self-reports, behavioral deferral, and dynamic tracking of trust over time~\cite{jian2000trust,desai2012dynamic}. Multi-dimensional trust models further distinguish perceived competence, predictability, transparency, and integrity~\cite{ashoori2019ai}, offering richer insight into how trust is formed and calibrated.

Task characteristics strongly influence reliance. Users tend to defer more to AI as task complexity increases, even when additional reliance is not warranted~\cite{bansal2019beyond,poursabzisangdeh2021manipulating}. Prognostic tasks, which involve long-term predictions, are particularly prone to miscalibrated reliance because users cannot easily verify outcomes~\cite{dearteaga2019evaluating}. This pattern is especially relevant in AD progression prediction, which is inherently uncertain and high-stakes. Understanding how users interpret uncertainty in such tasks is crucial for designing AI systems that promote appropriate reliance, rather than blind trust or excessive skepticism.

\subsection{Uncertainty Visualization in AI Systems}

Uncertainty visualization is a prominent strategy for helping users understand model limitations and calibrate reliance in complex decision-making tasks~\cite{padilla2020uncertainty, padilla2021uncertain}. Visual encodings such as color gradients, transparency, and numerical uncertainty labels provide perceptual cues about model confidence and ambiguity~\cite{boukhelifa2012evaluating,breslow2009cognitive}. These cues can shape user attention, influence perceived reliability, and alter behavioral reliance on AI recommendations~\cite{daradkeh2017incorporating,doula2022visualization,marusich2023using,cassenti2023representing}.

Studies show that task context and uncertainty type significantly shape how visualizations affect users. For example, salient uncertainty displays in hospital bed forecasting reduced users’ willingness to follow AI predictions~\cite{leffrang2021should}. Continuous uncertainty encodings increased trust and confidence in gaming scenarios, particularly among users initially skeptical of AI~\cite{reyes2025trusting}. Ordinal uncertainty expressions improved reliance calibration in predictive tasks such as college admissions forecasting~\cite{zhao2020uncertainty}. Individual differences such as numeracy, prior experience, and decision-making style further moderate these effects~\cite{kay2016ish}.

\textcolor{blue}{In AD research, a small number of prior works were found that highlight the importance of advanced visualization methods for addressing the complexities of AD biomarkers and their interaction dynamics. One of these studies introduced graph-based AI models to integrate multi-omic data to identify biomarker interactions essential for understanding AD progression~\cite{ zhang2024mosgraphgen, zhang2025mosgraphflow}. Other AI-driven studies investigate convolutional neural networks (CNNs) and their effectiveness in interpreting MRI data for AD-related biomarkers, demonstrating the value of computational approaches for visualizing biomarker relationships~\cite{wang2025early}. Notably, clinical research underscores the need for tools capable of integrating multiple clinical assessments to visualize AD biomarker networks and support reliable, patient-specific decision-making. Such tools are viewed as critical for developing a more comprehensive understanding of AD mechanisms~\cite{salimi2018can, guo2009model}. Given the limited research in this area, the literature points to the need for more advanced and integrative visualization frameworks, stronger support for long-term prognosis modeling and trajectory analysis as well as human-centered, explainable and responsible AI systems~\cite{el2023trustworthy, viswan2024explainable}.} Additionally, uncertainty visualization remains underexplored, particularly for long-term prognostic predictions. Eslami et al.~\cite{eslami2023unique} and Lai et al.~\cite{lai2023towards} have begun integrating uncertainty into AD prediction tools, highlighting the need for clearer communication of model limitations. However, prior work has not examined how different user groups, such as general participants and clinical experts, interpret and act upon visualized uncertainty in AD progression predictions.

\section{Objectives}
This section outlines the research gap motivating our work, the goals of the study, and the research questions and hypotheses that guide our investigation.

\subsection{Research Gap and Objectives}
\textcolor{blue}{Across clinical decision support systems (CDSS), human--AI trust research, and uncertainty visualization, prior work highlights the importance of calibrated reliance and the need to communicate AI limitations clearly. Yet we lack empirical understanding of how uncertainty visualizations influence trust and decision-making in \textit{long-term, high-uncertainty medical prognosis}, where outcomes unfold over months or years, uncertainty is intrinsic, and users rarely receive feedback to recalibrate trust. This gap is especially salient for AD progression prediction, where clinical decisions are high-stakes and model confidence intervals can be substantial.}

Our objective is therefore to examine how different uncertainty encodings shape users’ interpretations of AI-generated AD progression trajectories, and how these encodings influence trust, confidence, and decision-making. \textcolor{blue}{Moving beyond prior work that focuses primarily on classification or short-term diagnostic tasks, we study uncertainty in a prognostic scenario where ambiguity is unavoidable and interpretive demands are high.} Because uncertainty does not operate in isolation, we also examine how different levels of model transparency influence users’ interpretations and trust in prognostic outputs, an open question in clinical AI where stakeholders disagree on how much information is necessary for appropriate reliance. To understand how these effects manifest for different stakeholders, we study these uncertainty encodings separately with broadly recruited participants and with domain experts, allowing us to characterize distinct patterns in how each group interprets, evaluates, and relies on AI-generated prognostic information.

\subsection{Research Questions and Hypotheses}

Guided by this motivation, we investigate how uncertainty visualization and model information shape users’ interpretations of predicted CN/MCI/AD transitions, and how these factors influence trust, confidence, and decision-making. We address the following research questions and hypotheses.

\vspace{4pt}
\noindent\textbf{RQ1: Model information and trust.}  
\emph{How does the amount of information about the AI model (minimal vs.\ moderate) influence users' trust in AI-supported predictions of AD progression?}

\noindent\textbf{H1:} Providing \emph{moderate} model information will lead to \textcolor{blue}{\emph{more appropriately calibrated trust}} than providing minimal information.

\vspace{6pt}
\noindent\textbf{RQ2: Uncertainty visualization and trust.}  
\emph{How do binary vs.\ continuous uncertainty encodings affect users’ trust in AI predictions and their assessments of specific trust facets?}

\noindent\textbf{H2:} Continuous uncertainty visualizations will yield \textcolor{blue}{\emph{more appropriately calibrated in-task trust}} than binary visualizations.  
\noindent\textbf{H3:} Continuous uncertainty visualizations will yield \emph{higher ratings on specific trust facets}, particularly perceived \emph{reliability}, than binary visualizations.

\vspace{6pt}
\noindent\textbf{RQ3: Uncertainty, model information, and confidence.}  
\emph{How do uncertainty format and model information influence confidence in decisions based on AI predictions, and how do these effects differ when examined separately with general users and domain experts?}

\noindent Confidence analyses were exploratory due to mixed evidence in prior work on whether uncertainty increases or decreases decisional confidence.

\vspace{6pt}
\noindent\textbf{RQ4: Expert trust and decision-making.}  
\emph{How do domain experts’ trust, confidence, and decision judgments vary when uncertainty is visualized continuously versus in binary form?}

\noindent\textbf{H4:} Continuous uncertainty visualizations will yield \textcolor{blue}{ \emph{more appropriately calibrated trust}} among domain experts, both at the trial level and across multi-dimensional trust facets.

\section{Methodology}

\subsection{Data and Model Setup}


We implemented an instance of the Machine Learning for Visualizing AD (ML4VisAD) model~\cite{eslami2023unique} to convert multimodal input data into color-coded predictions that represent disease progression over time. The ML4VisAD model’s architecture consists of two main parts: the feature extraction phase and the image production phase. The first part of the network focuses on extracting features from various modalities (MRI, PET, etc.) and cognitive assessments through fully connected layers and concatenation of features. The second part of the model involves tensorization, a process that converts feature representations into appropriately shaped tensors, followed by reshaping these tensors into 2D formats and applying transposed convolutional operations to reconstruct a 2D image from the extracted features. The architecture of the ML4VisAD model is available in a public repository~\footnote{\url{https://github.com/mohaEs/ML4VisAD}}.

\subsection{Stimuli: AI-Generated AD Trajectories}

\textcolor{blue}{Central to this study are the visual stimuli generated by the ML4VisAD model. We selected this visualization scheme because of its novel estimator design, in which multimodal input features are encoded as a 2D RGB image representing AD progression. During the image-generation phase, the model produces a $23 \times 23 \times 3$ image that is compared with a ground-truth image of the same dimensions. Before training, ground-truth images are generated based on the clinical statuses at each visit recorded in the ADNI QT-PAD database~\footnote{\url{https://ida.loni.usc.edu}}. These images encode the patient’s disease stage using color-coded segments. During training, the model learns to minimize the error between its predicted images and these ground-truth representations.}

\textcolor{blue}{The dimensions of ground-truth images define the spatial layout, or template, through which the model learns feature interactions during training. This template consists of five segments. The first four correspond to disease stages, each spanning 5 pixels along the \textbf{horizontal axis} and representing a clinical time point (baseline = 0, 6 months, 12 months, 24 months). The fifth segment has a distinct semantic role: it encodes the degree of uncertainty that the ML4VisAD model may introduce during feature extraction and multimodal data visualization. In this segment, lower pixel intensities indicate higher uncertainty, signaling that the model’s predictions reflect ambiguity in mapping the input features to the specific region being modeled. The \textbf{vertical axis} reflects one spatial dimension of the 2D tensor; it does not encode clinical information but instead serves as a spatial structure that enables the transposed convolutional layers to learn visual patterns. Both the x- and y-axes describe how multimodal signals are organized in 2D space to produce the final color-coded representation of disease trajectory. By default, the image dimensions are set to $23 \times 23 $ pixels, though increasing it to $45 \times 45$ pixels would produce smoother transitions~\cite{eslami2023unique}. We chose to keep the dimensions at the default value to better categorize images by the level of uncertainty reflected in their pixel values.} Colors indicate the predicted diagnosis at each visit (green = cognitively normal, blue = mild cognitive impairment, red = AD). Images with a single color across all visits represent stable cases (e.g., stable CN, stable MCI, stable AD), while images with two or more colors indicate progression or conversion (e.g., CN$\rightarrow$MCI, MCI$\rightarrow$AD; Figure ~\ref{fig:stimulus}). 

\begin{figure}[h]
\centering
\includegraphics[width=\linewidth]{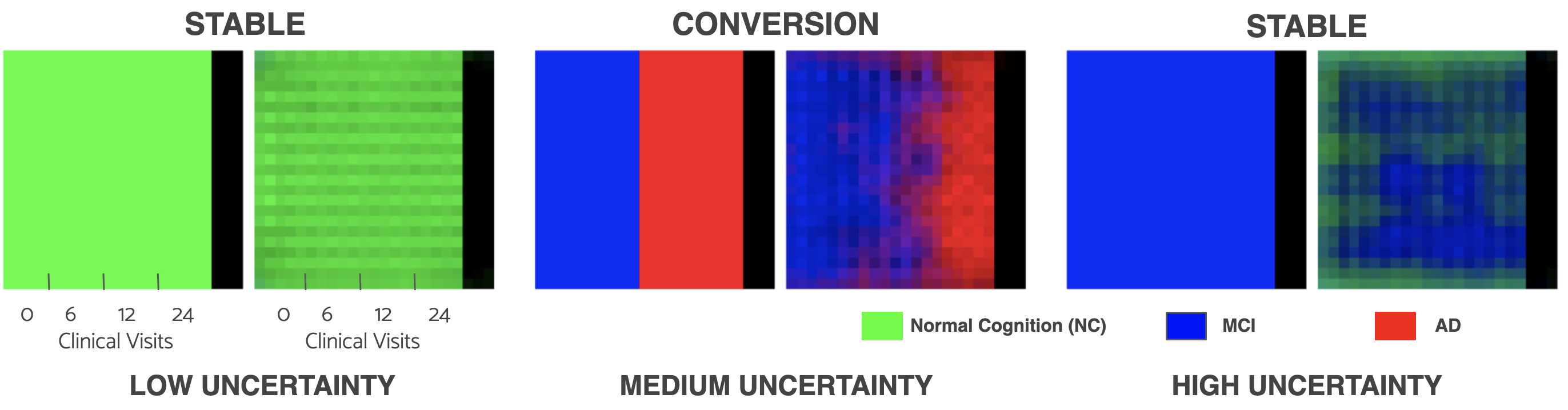}
\caption{Example stimuli. Each row shows a case (stable CN, converting MCI$\rightarrow$AD, stable MCI). Columns show binary (left) vs.\ continuous (right) uncertainty formats.}
\label{fig:stimulus}
\end{figure}

\paragraph{Uncertainty visualization formats.}
We manipulated \emph{how} uncertainty was visualized while keeping predictions constant:
\begin{itemize}
    \item \textbf{Binary uncertainty}: the right-most three pixel rows served as an indicator strip. Filled (black) rows denoted high uncertainty; empty rows denoted low uncertainty.
    \item \textbf{Continuous uncertainty}: as well as the right-most three pixel indicator strip, uncertainty was also encoded through color saturation. Images were converted to CIELAB space, and luminance ($L$) was used as a certainty proxy—higher $L$ produced more vivid colors (higher certainty), and lower $L$ produced darker, less saturated colors (higher uncertainty)~\cite{seymour2020does}.
\end{itemize}

In all conditions, uncertainty was always shown; the manipulation focused on \emph{how} it was represented.

\begin{figure}[h]
\centering
\includegraphics[width=0.75\linewidth]{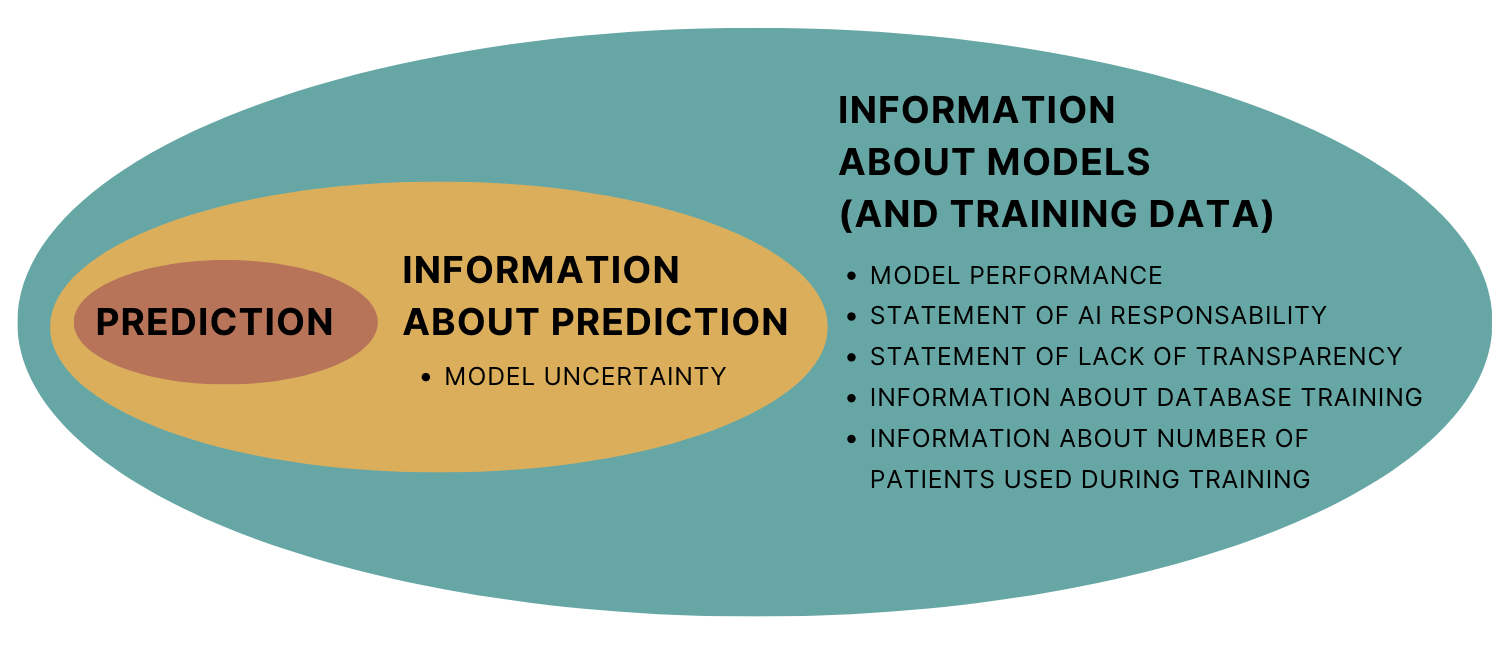}
\caption{Information elements shown to participants in the ``moderate information’’ condition, adapted from Lai~\textit{et al.}~\cite{lai2023towards}.}
\label{fig:ai_assistance_elements}
\end{figure}

\section{User Studies}
\textcolor{blue}{We conducted two user studies to examine how uncertainty visualizations influence trust, confidence, and decision-making in AI-supported predictions of AD progression. Study~1, a mixed-method online study with a broad participant sample, investigated how general users interpret AI-generated trajectories under different uncertainty encodings. Study~1 also manipulated a second factor: the amount of information provided about the AI model and its training process (minimal vs.\ moderate). Insights from Study~1 informed revisions to the design of the second study. Study~2 focused on domain experts in neurology, neuroimaging, and dementia research, evaluating how trained professionals interpret the same uncertainty encodings. To isolate the effect of uncertainty visualization while respecting experts’ limited availability, Study~2 did not vary the level of model information; instead, all experts received a standardized \emph{moderate-information} description of the model, including dataset details, cohort size, and a simplified summary of the training procedure.}

\textcolor{blue}{Across both studies, dependent measures included trial-level trust in the AI predictions, decision confidence, and, in Study~2, whether experts’ decisions changed when viewing the same patient case with different uncertainty encodings. A pilot study preceded the main experiments to refine the task interface, tutorial, and model-information materials.} All studies were approved by our institution’s research ethics board.

\subsection{Pilot Study}
\textcolor{blue}{We first conducted a pilot experimental validation study with 5 participants, independent from the recruited cohort that participated in Study~1. Insights from this pilot informed revisions to the design of User Study 1. Using a Think-aloud protocol, we identified multiple usability challenges, including: (1) the need to revise the information presented on the tutorial page, (2) consolidating the tutorial from multiple pages into a single page, (3) ensuring uniformity across trust scales by standardizing all scales to 10 points, and (4) adjusting the amount of information provided in the ``moderate information'' condition by adding details about the model’s training process, and (5) ensuring optimal selection of images that expressed uncertainty in different levels: low, medium, and high.
}

\subsection{Study 1: Broad Participant Study}
\textcolor{blue}{We used a 2 (AI model information: minimal vs. moderate; between-subjects) × 2 (uncertainty format: continuous vs. binary; within-subjects) mixed design, in which all participants saw both types of uncertainty visualization but were randomly assigned to one of the two AI-information conditions.} \textcolor{blue}{A priori statistical power analysis for a between-subject comparison assuming a large effect size ($d \approx 0.8$), a significance level of $\alpha = .05$, and 80\% statistical power, indicated that approximately 42 participants (21 per group) would be needed.} Participants were recruited through institutional mailing lists and social media platforms (LinkedIn and X). The study followed an online workflow consisting of informed consent, a tutorial, a pre-test questionnaire, a main prediction task, and a post-test questionnaire. Participants were randomly assigned to one of two information conditions describing the AI model: \textit{minimal detail} (no model background) or \textit{moderate detail} (dataset description, training cohort size, inclusion/exclusion criteria, and a note that hyper-parameters were available upon request).

\subsubsection{Tutorial}
\textcolor{blue}{The tutorial introduced users to AD and the importance of early detection, then guided them through examples of AI-generated predictions showing how the disease may remain stable or progress over time. Users viewed visualizations of stable stages: cognitive normal (CN), mild cognitive impairment (MCI), and Alzheimer’s disease (AD), along with binary uncertainty indicators highlighting where the model lacked confidence. They then explored converting stages, such as CN $\rightarrow$ MCI or MCI $\rightarrow$ AD, using continuous uncertainty visualizations encoded through color saturation. Finally, the tutorial demonstrated how human--centered design principles can enhance the interpretability of AI predictions by helping clinicians distinguish between high-certainty and high-uncertainty outputs when confirming or validating an AD prognosis.}

\subsubsection{Pre-test Questionnaire}
Participants provided demographic information and rated their familiarity with AI development, data visualization, and AD on 10-point scales.

\subsubsection{Task: AI Uncertainty Predictions for AD Progression}
Participants viewed 12 AI-generated predictions of patient trajectories, for both stable and progressive cases. Each participant saw six stimuli with binary uncertainty encoding and six with continuous encoding, \textcolor{blue}{presented in a blocked between-subject design with randomized block order, with the amount of information provided about the AI model and
its training process as the independent variable.} For each stimulus, participants judged whether the patient remained stable or transitioned (e.g., CN\,$\rightarrow$\,MCI, MCI\,$\rightarrow$\,AD) at 6, 12, or 24 months. Stimuli spanned three uncertainty levels: low (0--10\%), medium (11--20\%), and high ($\geq$21\%).

\subsubsection{Measures}
After each stimulus, participants reported (1) the perceived transition point, (2) decision confidence (10-point scale), and (3) trust in the human--AI decision-making process (10-point scale). After each block, they completed the adapted Ashoori and Weisz's~\cite{ashoori2019ai} trust scale, assessing trustworthiness, reliability, technical competence, and personal attachment.

\subsubsection{Post-test Questionnaire}
Participants provided open-ended reflections on how uncertainty influenced their decisions, how they interpreted the visualizations, and what factors shaped their trust in the AI predictions.

\subsection{Study 2: Expert Study}
This study engaged domain experts in neurology, neuroimaging, and dementia research (AI researchers working in AD) to assess how experienced practitioners interpret uncertainty in AI-based AD progression predictions. \textcolor{blue}{Experts were recruited using a combination of targeted email outreach and \textit{snowball sampling}: initial contacts were invited to participate and encouraged to share the study with qualified colleagues. Snowball sampling is commonly used in medical AI research when the eligible population is small and highly specialized. Prior to recruitment, an a priori power analysis for a within-subject comparison using a large effect size (based on results from Study~1) ($d \approx 1.1$), $\alpha = .05$, and 80\% power indicated that approximately 8--10 experts would be sufficient. We therefore targeted 8--12 participants and successfully recruited $N = 10$ experts.}

Procedures mirrored those of Study~1, with consent, a pre-test questionnaire, a main task, and a post-test questionnaire. Experts also received background information about the AI model, including dataset details, cohort size, inclusion/exclusion criteria, and a simplified description of the training procedure.

\subsubsection{Pre-test Questionnaire}
Experts reported demographic information and rated their knowledge of AI development, data visualization, and AD on 10-point scales.

\subsubsection{Task: AI Uncertainty Predictions for AD Progression}
\textcolor{blue}{To respect experts’ limited availability, we streamlined the task by reducing the total number of stimuli. Each expert viewed six patient trajectories, each presented twice, once with binary uncertainty and once with continuous uncertainty, for a total of 12 stimuli.} Uncertainty was categorized into two levels, low (<30\%) and high (>30\%). Experts completed the same core judgment task as in Study~1, indicating whether each patient remained stable or transitioned at 6, 12, or 24 months. Stimuli were shown in a blocked within-subjects design with randomized order of uncertainty format.

\subsubsection{Measures}
The same measures as in Study~1 were collected: transition point, decision confidence, and trust ratings for each stimulus.

\subsubsection{Post-test Questionnaire}
Experts reflected on whether uncertainty visualizations improved interpretability, clarified model limitations, conveyed reliability, or supported potential adoption in research or clinical practice. Open-ended responses captured detailed reasoning about visualization strengths and limitations.

\section{Results}

\subsection{Broad Participant Study (Study 1)}
\textcolor{blue}{We collected responses from 43 participants, with 21 randomly assigned to the minimal-information condition and 22 to the moderate-information condition. However, several responses were incomplete and had to be excluded. In total, we analyzed data from 37 participants, 19 in the first group and 18 in the second} (59\% male, 41\% female; $M_{age}$ = 27.8, range = 21--38), primarily graduate students across VR/AR, HCI, ML/DL, neuroimaging, psychology, computer vision, photonics, and formal methods. \textcolor{blue}{Participants represented multiple nationalities (Canada, United States, Mexico, Iran, China, Turkey), which provided a diversity of perspectives and helped reduce cultural bias in interpreting the visualizations}. On 10-point self-report scales, participants rated their experience with AI development ($M$ = 6.1, SD = 2.63), AI usage ($M$ = 7.3, SD = 1.77), data visualization ($M$ = 6.1, SD = 2.09), and AD ($M$ = 3.1, SD = 2.33).\footnote{All statistical analyses were conducted in JASP v.0.17.2.1, https://jasp-stats.org}

\subsubsection{Task 1: AD Progression}
We first assessed the effect of AI model information (Figure~\ref{fig:trust_info}). Trust scores were slightly higher with moderate information ($M$ = 6.99, SD = 1.64) than with minimal information ($M$ = 6.44, SD = 1.66), but this difference was not significant in either independent $t$-tests or linear mixed models. Participants who received minimal model information nevertheless expressed a desire for more context: \textit{``I'd trust the predictions more if I know more about how they are produced and visualized.''} This finding fails to support H1, although qualitative evidence suggests that richer or layered model explanations may be required to meaningfully influence trust.

\subsubsection*{Trust}
To test H2, we examined participants’ in-test ratings of trust in AI predictions, reported after each trial. Binary encodings of uncertainty were rated significantly higher in trust ($M$ = 7.19, SD = 2.52) than continuous ones ($M$ = 6.22, SD = 1.36), $t(36) = 2.550, p < .05$. A linear mixed model that accounted for repeated measures confirmed this effect ($p < .05$). Participants explained that while continuous formats conveyed richer information, they often appeared ``blurred'' or ``noisy,'' making decision boundaries less clear and reducing trust: \textit{``Despite providing more information, the blurred transition made the prediction vague and reduced my trust.''} These results provide only partial support for H2: using continuous uncertainty did not increase immediate trust, as binary encodings were trusted more in the moment.  

We also assessed block-level ratings of the four facets of trust~\cite{ashoori2019ai}, collected after participants completed each visualization condition (Figure~\ref{fig:facets}). Continuous uncertainty visualizations significantly increased perceived \emph{Reliability} ($M$ = 6.92, SD = 1.99) compared to binary encodings ($M$ = 4.87, SD = 2.83), $t(36) = -3.768, p < .001$. No significant differences emerged for the other facets. Qualitative feedback echoed this distinction: \textit{``If there is no uncertainty visualized [binary uncertainty], it is easy to be overly confident in the model. The more it is visualized, the more I can be trustworthy of the decisions.''} Binary encodings were also described \textit{``uninformative and hard to trust.''} These results support H3: continuous encodings enhanced perceived reliability, even though they reduce trust ratings in the moment of making the decision.

\begin{figure}[h]
\centering
\begin{minipage}{0.4\linewidth}
    \centering
    \includegraphics[width=\linewidth]{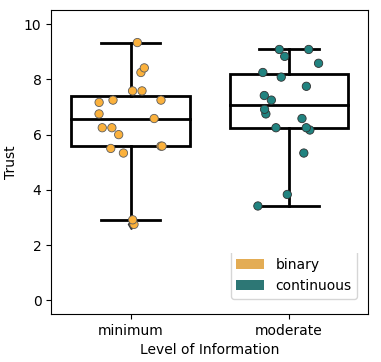}
    \caption{Effect of model information levels on trust.}
    \label{fig:trust_info}
\end{minipage}
\hfill
\begin{minipage}{0.4\linewidth}
    \centering
    \includegraphics[width=\linewidth]{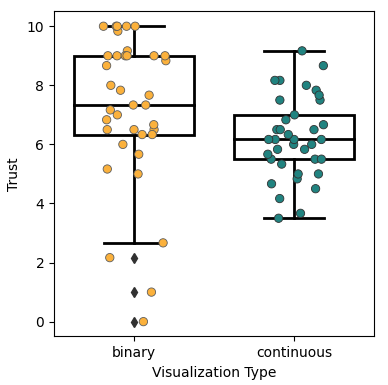}
    \caption{Comparison of binary versus continuous uncertainty encodings and their effects on trust.}
    \label{fig:trust_visual_type}
\end{minipage}
\end{figure}

\begin{figure*}
\centering
\includegraphics[width=0.9\linewidth]{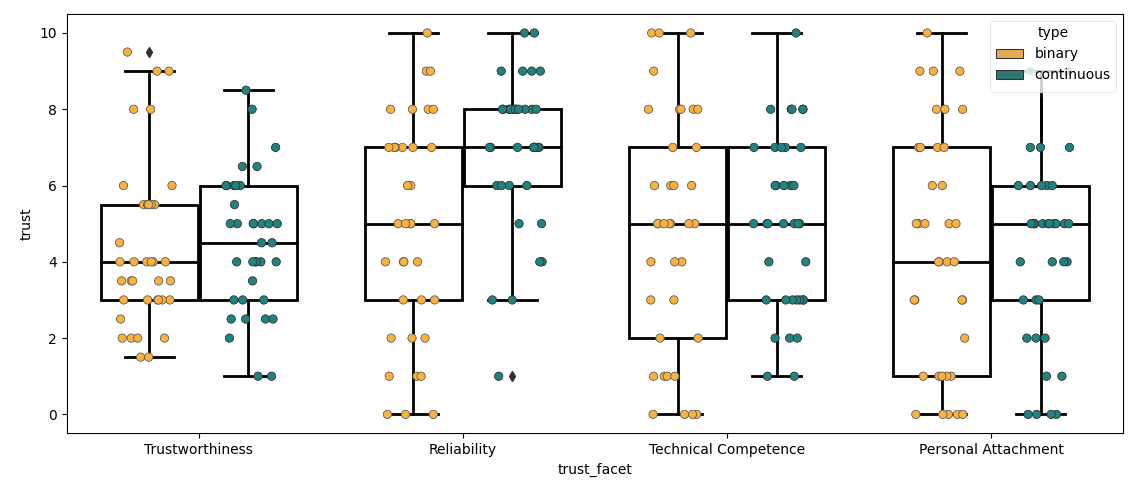}
\caption{Comparison of visual representations of AI uncertainty (binary vs continuous) using the multi-dimensional scheme of trust according to Ashoori and Weisz~\cite{ashoori2019ai}.} \label{fig:facets}
\end{figure*}

As observed, these findings show that participants distinguished between \emph{overall trust in the moment} and \emph{specific trust facets assessed afterward}. Continuous encodings improved reliability judgments (supporting H3) but reduced overall trust (contradicting H2), whereas binary encodings preserved higher global trust but offered weaker signals of reliability. The weak effect of model information level suggests that H1 was not supported in this study. 

\subsubsection*{Confidence}
To address RQ3, we examined participants’ confidence ratings, collected after each trial. As shown in Figure~\ref{fig:confidence_info}, the level of model information significantly influenced confidence: participants reported lower confidence with minimal detail ($M$ = 6.83, SD = 2.42) compared to moderate information ($M$ = 8.70, SD = 1.54), $F(35) = 9.953, p < .01$. Thus, providing at least a moderate level of model detail increased participants’ confidence in their decisions based on AI model output. Additionally, the type of encoding of AI uncertainty also had a significant effect on confidence (Figure~\ref{fig:confidence_encodings}). Binary uncertainty visualizations elicited higher confidence ($M$ = 8.70, SD = 1.34) than continuous visualizations ($M$ = 6.83, SD = 1.36), $t(36) = 8.128, p < .001$. A linear mixed model confirmed this difference after accounting for repeated measures and inter-individual variability ($p < .001$). These results provide a clear answer to RQ3: participants felt more confident with binary encodings than with continuous ones.

\begin{figure}[h]
\centering
\begin{minipage}{0.5\linewidth}
    \centering
    \includegraphics[width=\linewidth]{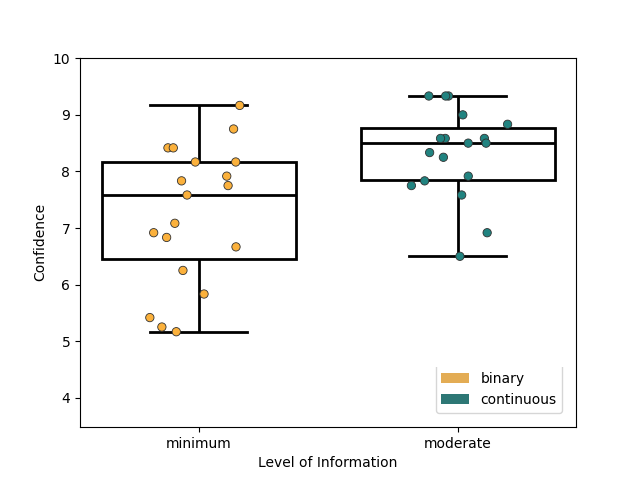}
    \caption{Effect of model information levels on confidence.}
    \label{fig:confidence_info}
\end{minipage}
\hfill
\begin{minipage}{0.45\linewidth}
    \centering
    \includegraphics[width=\linewidth]{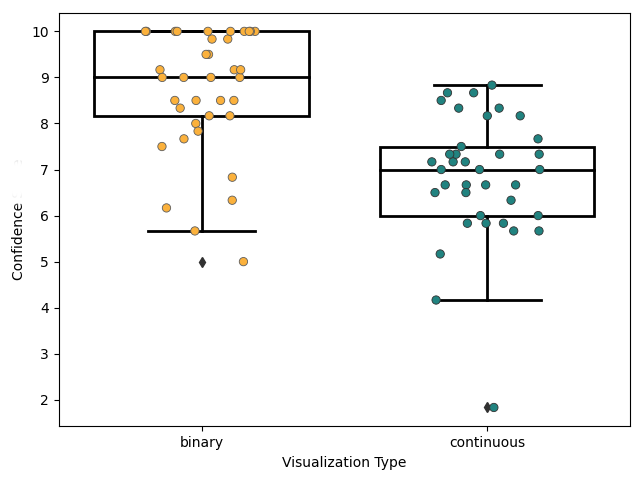}
    \caption{Comparison of binary versus continuous uncertainty encodings and their effects on confidence.}
    \label{fig:confidence_encodings}
\end{minipage}
\end{figure}

Qualitative feedback adds nuance to these findings. Several participants reported that seeing uncertainty information increased their confidence because it helped them gauge how much to rely on the model: \textit{``If I did not know the uncertainty I wouldn’t trust my decision at all''}. Another echoed \textit{``I liked seeing uncertainty, it made me realize how much I can (and should) trust the data''}. At the same time, others felt that when uncertainty was displayed more continuously it sometimes undermined confidence, making choices less clear \textit{``Visualizing uncertainty did slow the decision making and reduced the confidence but it was more informative''}.  

Taken together, these results suggest that while uncertainty visualization can boost confidence, the effect depends on its form. At first glance, it may seem counterintuitive that binary encodings produced higher confidence, since continuous encodings convey richer detail about model certainty and might be expected to support more informed decisions. However, qualitative feedback suggests the opposite. The additional visual noise and blurred boundaries in continuous encodings sometimes introduced doubt rather than clarity. As one participant noted, continuous formats made decision boundaries ``vague'' lowering their confidence in their decision. In contrast, binary encodings, though coarser, provided clearer categorical cues that supported decisional assurance, even at the cost of obscuring model uncertainty.

\subsection{Expert Study}
We recruited 10 domain experts with more than 10 years of experience in neuroimaging and dementia research (AI researchers working in dementia), including researchers specializing in MRI, aging, and AI, as well as professors, neurologists, and data scientists. Participants’ mean age was 42.8 years (range: 35–51); 8 identified as male and 2 as female. On a 1 (no experience) to 10 (very experienced) scale, participants reported high expertise in AI development ($M=8.0$, $SD=2.16$), AI usage ($M=8.3$, $SD=1.82$), data visualization ($M=7.4$, $SD=1.50$), and AD ($M=6.9$, $SD=2.55$).

\subsubsection{Task: AI Uncertainty Predictions for AD Progression}

\subsubsection*{Trust}
Figure~\ref{fig:expert_task_trust} (left) shows experts’ trust ratings across tasks, comparing AI predictions with and without visualized uncertainty, split by low (<30\%) and high (>30\%) uncertainty conditions. Trust was generally higher for continuous visualizations than for binary ones, particularly under high uncertainty. Paired-sample $t$-tests confirmed that continuous encodings significantly increased trust in model predictions for high-uncertainty trials ($p < .05$), suggesting that visualizing uncertainty in a continuous form can make model limitations more transparent. Qualitative feedback echoed this pattern: experts noted that continuous displays helped them better gauge prediction reliability.

Block-level ratings of the four facets of trust emphasized the benefits of continuous uncertainty visualizations. As shown in Figure~\ref{fig:expert_post_trust1}, trust ratings were markedly higher across all facets when uncertainty was displayed continuously. In the binary condition, mean scores were low: Personal Attachment ($M=1.8$, $SD=2.35$), Reliability ($M=2.6$, $SD=3.31$), Technical Competence ($M=1.6$, $SD=1.84$), and Trustworthiness ($M=1.8$, $SD=2.04$). In contrast, the continuous condition yielded substantially higher ratings: Personal Attachment ($M=8.2$, $SD=2.35$), Reliability ($M=7.4$, $SD=3.31$), Technical Competence ($M=8.4$, $SD=1.84$), and Trustworthiness ($M=8.2$, $SD=2.04$). Paired-sample $t$-tests confirmed statistically significant differences for all trust facets (Trustworthiness $p < .001$, Reliability $p = .047$, Technical Competence $p < .001$, Personal Attachment $p < .001$). As such, these findings provide strong support for \textbf{H4}, demonstrating that continuous uncertainty visualizations significantly enhance experts’ trust in AI predictions.

\begin{figure*}[t]
\centering

\begin{minipage}[b]{0.48\linewidth}
    \centering
    \includegraphics[width=\linewidth]{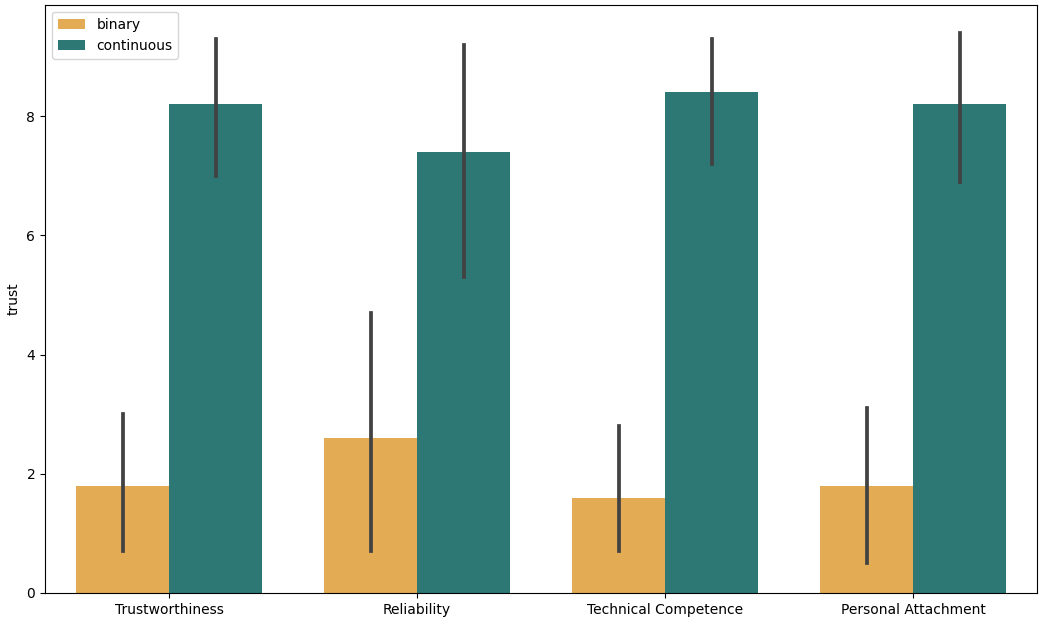}
    \caption{Comparison of binary and continuous AI uncertainty visualizations using Ashoori and Weisz's multi-dimensional trust framework~\cite{ashoori2019ai} among experts.}
    \label{fig:expert_post_trust1}
\end{minipage}
\hfill
\begin{minipage}[b]{0.48\linewidth}
    \centering
    \includegraphics[width=\linewidth]{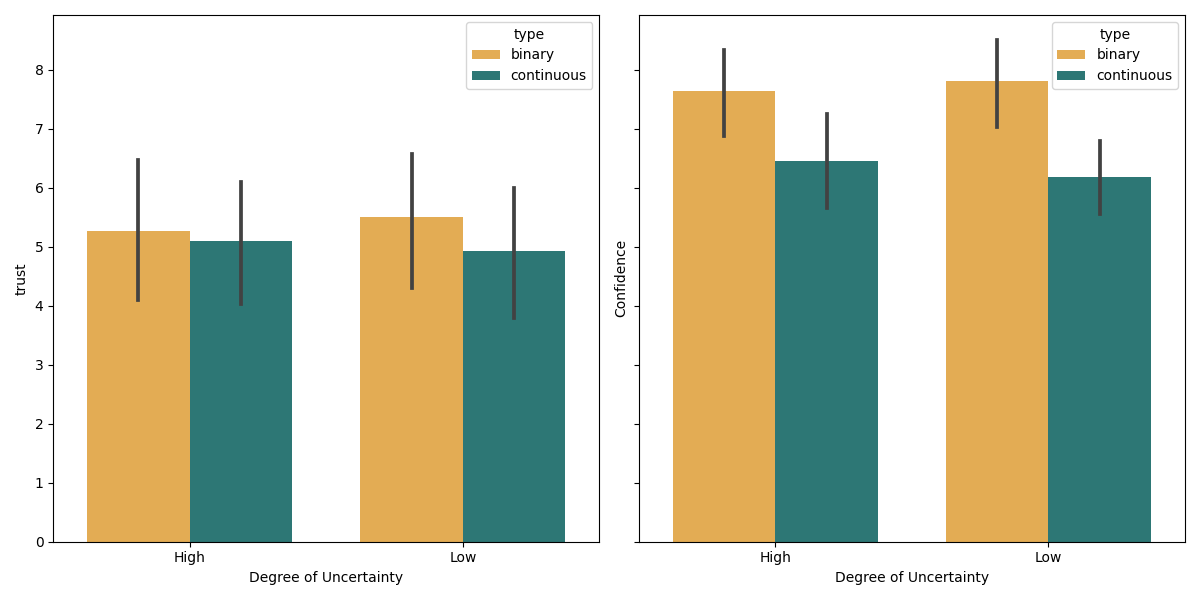}
    \caption{Trust \textit{(left)} and confidence \textit{(right)} distributions among experts by visualization type and uncertainty level.}
    \label{fig:expert_task_trust}
\end{minipage}

\end{figure*}

\subsubsection*{Confidence}
Figure~\ref{fig:expert_task_trust} (right) shows experts’ confidence ratings in their decisions. Confidence was significantly higher with binary visualizations for both high-uncertainty ($p = .014$) and low-uncertainty trials ($p = .003$). These results suggest that binary outputs may foster overconfidence by presenting predictions as categorical, whereas continuous visualizations provide more nuanced cues that encourage calibrated and cautious decision-making.

\subsubsection*{Decision Change}  
A \emph{decision change} was defined as a difference in judgment for the same patient when predictions were shown with binary versus continuous visualizations. Experts were not informed that the same patient case appeared twice, ensuring that any decision differences reflected the visualization format rather than deliberate cross-checking. Across all trials, experts changed their decisions in 23 of 66 cases (34.8\%), while the remaining 65.2\% of responses were consistent across formats. When broken down by uncertainty level, decision changes occurred 12 times under high-uncertainty conditions (36.4\%) and 11 times under low-uncertainty conditions (33.3\%). A chi-square test indicated that this distribution was not significant ($\chi^2 = 0.0$, $p = 1.0$). When examining the interaction between uncertainty level and visualization mode, decision changes were more frequent under high-uncertainty conditions (17 of 48 trials; 35.4\%) than under low-uncertainty conditions (10 of 48 trials; 20.8\%). Although this suggests that continuous visualizations paired with high uncertainty may have prompted more reconsideration of judgments, a chi-square test indicated that the difference was not statistically significant ($\chi^2 = 1.86$, $p = .17$).

Among the 23 changes, 14 (60.9\%) were with continuous visualizations and 9 (39.1\%) were with binary encodings. Although this suggests that experts more often revised their judgments when viewing continuous visualizations, a binomial test confirmed that this difference was not statistically significant ($p = .40$).

\begin{table}[h]
\centering
\caption{Decision changes by visualization direction and uncertainty level.}
\resizebox{\columnwidth}{!}{%
\begin{tabular}{lccc}
\toprule
 & \textbf{High Uncertainty} & \textbf{Low Uncertainty} & \textbf{Total} \\
\midrule
No change & 21 & 22 & 43 \\
Binary $<$ Continuous & 3 & 11 & 14 \\
Binary $>$ Continuous & 9 & 0 & 9 \\
\midrule
\textbf{Total} & 33 & 33 & 66 \\
\bottomrule
\end{tabular}%
}
\end{table}

\subsubsection*{Post-test Questionnaire}
Experts reported needing less supplementary information about the AI model when uncertainty was visualized continuously ($M=7.5$, $SD=1.62$) compared to binary encodings ($M=8.8$, $SD=7.5$). Open-ended feedback echoed this pattern: continuous uncertainty displays were described as enhancing the credibility of AI predictions and supporting interpretation of varying confidence levels.

At the same time, experts consistently emphasized that uncertainty alone was insufficient for validation. Trust was more strongly grounded in external validation, transparency, and demonstrated model performance. As one AI neuroimaging expert explained \textit{``Uncertainty is not a proof of validity''} and \textit{``evaluation on external datasets would be a better way to increase my trust in AI predictions''}. Others stressed the value of human--centered explanations, suggesting that ``human--like explanations'' would be more effective in fostering trust than probabilistic visualizations alone.

While uncertainty visualizations added some credibility, by explicitly acknowledging prediction uncertainty and differentiating between high- and low-probability outcomes, this credibility was conditional. Experts highlighted that visualizations were meaningful only when paired with relevant information about model training, assumptions, and limitations: \textit{``It’s nice to visualize uncertainty, but it doesn’t mean much to me unless I get more information about the model training process''}. Overall, the feedback suggests that uncertainty visualizations are useful but must be embedded within a broader framework of transparency and interpretability. Without sufficient context and validation, their ability to meaningfully shape trust, particularly among experts, may be limited.

\subsubsection*{Synthesis of findings}
In Study 2, continuous uncertainty visualizations increased experts’ trust, both in trial-level ratings and across all post-task trust facets, while binary visualizations yielded higher confidence, pointing to a potential risk of overconfidence. Experts changed their decisions in about one-third of cases when viewing the same patient with different encodings, though change rates did not vary by uncertainty level or visualization mode. Post-task feedback indicated that continuous visualizations enhanced credibility and reduced the need for supplementary model information, but experts emphasized that trust also depended on access to details about model training and evaluation.

\subsection{Summary}
In summary, the two studies show that visualization of AI uncertainty formats and user expertise jointly shape trust and confidence in AI predictions. Generalist participants trusted binary encodings more in the moment but judged continuous formats as more reliable, whereas experts showed consistently higher trust across all facets with continuous visualizations, even though binary encodings inflated their confidence. Table~\ref{tab:results-summary} summarizes these findings, highlighting how expertise moderates the trade-offs between clarity, credibility, and overconfidence. ~\textcolor{blue}{These contrasts not only suggest a need for adaptive visualization strategies that account for different levels of expertise, but also show that uncertainty visualization, used as an XAI technique, can clarify an AI’s predictive capabilities, helping users recognize model limitations and calibrate their reliance when making complex decisions.}
\begin{table*}[h]
\centering
\scriptsize
\caption{Summary of results across Study 1 (Broad Participants) and Study 2 (Experts). Arrows indicate direction of higher scores/effects.}
\label{tab:results-summary}
\begin{tabular}{@{}p{2.5cm}p{4cm}p{7.5cm}p{1.5cm}@{}}
\toprule
\textbf{Outcome} & \textbf{Manipulation / Comparison} & \textbf{Main Finding} & \textbf{Support} \\
\midrule
\multicolumn{4}{@{}l}{\textbf{Study 1: Broad Participant Study}}\\
\midrule
Trust (info level) & Minimal vs.\ Moderate model info &
No significant difference; participants expressed desire for more context. &
H1: \xmark \\
Trust (per-trial) & Binary vs.\ Continuous uncertainty &
Binary $\uparrow$ trust vs.\ continuous (in-task). &
H2: \xmark \\
Trust facets (post-block) & Binary vs.\ Continuous uncertainty &
Reliability $\uparrow$ with continuous (other facets n.s.). &
H3: \checkmark \\
Confidence (info level) & Minimal vs.\ Moderate model info &
Moderate info $\uparrow$ confidence. &
RQ3: Moderate \\
Confidence (vis type) & Binary vs.\ Continuous uncertainty &
Binary $\uparrow$ confidence vs.\ continuous. &
RQ3: Binary \\
\midrule
\multicolumn{4}{@{}l}{\textbf{Study 2: Expert Study}}\\
\midrule
Trust (per trial) & Binary vs.\ Continuous uncertainty; Low vs.\ High uncertainty &
Continuous $\uparrow$ trust, especially under high uncertainty. &
H4: \checkmark \\
Trust facets (post-task) & Binary vs.\ Continuous uncertainty &
All trust facets $\uparrow$ with continuous (large differences). &
H4: \checkmark \\
Confidence & Binary vs.\ Continuous; Low vs.\ High uncertainty &
Binary $\uparrow$ confidence for both uncertainty levels (potential overconfidence). &
RQ3: Binary \\
Decision change & Same patient, binary vs.\ continuous uncertainty &
23/66 (34.8\%) changed; no significant differences by uncertainty level or visualization mode. &
RQ4: Exploratory \\
Model info need (post) & Binary vs.\ Continuous uncertainty &
Less supplementary model info reported as needed with continuous; experts still requested training/evaluation details. &
RQ4: Exploratory \\
\bottomrule
\end{tabular}

\vspace{4pt}
\raggedright \emph{Notes:} Checks (\checkmark) and crosses (\xmark) indicate hypothesis outcomes (supported / not supported). Arrows denote higher values. n.s.\ = not significant. RQ3 and RQ4 mark exploratory analyses related to confidence, decision change, and perceived need for additional model information.
\end{table*}

\section{Discussion}
Building on recent work on AI uncertainty estimation in AD prognosis~\cite{eslami2023unique}, our mixed-method study examined how generalist users and domain experts interpret and trust AI predictions when uncertainty is visualized in different formats. Across two populations, we compared binary (color/no color) and continuous (color saturation) encodings, as well as multiple visualization styles, to assess their influence on trust, confidence, interpretability, and decision stability. Our findings extend prior HCI research on uncertainty visualization and trust in AI by highlighting how expertise moderates the balance between clarity and transparency.  

Our results resonate with prior findings that interpretability alone does not always increase trust~\cite{poursabzisangdeh2021manipulating}, while outcome feedback and contextual cues often matter more~\cite{dietvorst2015algorithm}. In our studies, generalists trusted binary encodings more in the moment and reported higher confidence, but continuous visualizations improved reliability judgments and reduced the risk of overconfidence. \textcolor{blue}{This divergence between in-task trust and post-block reliability ratings suggests that participants may have mapped ``trust'' during the trial-level task to their perceived ability to make a decision rather than to a broader assessment of model quality. This aligns with prior work showing that users often conflate trust with perceived task ease or decisional clarity when interacting with decision-support systems~\cite{lee2004trust,dzindolet2003role}.} Experts, by contrast, showed consistently higher trust across all facets with continuous displays, even though binary formats inflated confidence. This divergence echoes arguments that trust is multi-dimensional~\cite{ashoori2019ai} and that user expertise shapes how uncertainty is perceived~\cite{wang2021addressing}.  

Participants in both studies emphasized that uncertainty visualization alone was insufficient for validating AI predictions. Consistent with trust theory~\cite{mayer1995integrative}, participants expected richer transparency about how models were trained and validated. Experts in particular requested details about datasets, inclusion/exclusion criteria, and external validation, reinforcing earlier HCI findings that explanations are more effective when paired with contextual evidence~\cite{bansal2021does}. Our results therefore extend prior work by showing that continuous uncertainty can reveal model limitations, but trust is strengthened only when such visual cues are integrated with human-readable model information.  

\subsection{Limitations and Future Work}  
Several limitations should be acknowledged. The generalist sample consisted primarily of graduate students with relatively high technical literacy, limiting generalizability to lay populations. \textcolor{blue}{This skew likely reflects network effects in our recruitment for the first user study. However, because this was an exploratory study focused on evaluating prototype usability, this group provided appropriate initial feedback. We ackowledge that online recruiting platforms are useful to find more and diverse participants. While these platforms can provide access to large and diverse participant pools, we avoided them due to well-documented challenges~\cite{wazny2017crowdsourcing}. Tasks like those in our study, which do not have a single correct answer, are particularly susceptible to data-quality issues, including careless or malicious responses. Additionally, these platforms make it difficult to ensure attentive and thoughtful engagement with the study materials. For these reasons, we opted for a more controlled recruitment approach, which, while less scalable, was better suited to the goals of our exploratory evaluation.} 

\textcolor{blue}{We acknowledge that the final sample sample size for the generalist study (N = 37)  was smaller than our a priori target (N = 42). This was due to several incomplete responses that had to be excluded. However, a sensitivity analysis conducted with the G*Power software revealed that, with our final sample size of $N = 37$, $\alpha = .05$, and $80\%$ power, the study was adequately powered to detect a minimum effect size of $d = 0.83$ (Cohen’s d).} The expert study was small (N=10) and streamlined to 12 stimuli, which improved feasibility but constrained statistical power and prevented exploration of medium levels of uncertainty. Finally, while we examined visualization format, we did not test interactions with outcome feedback, which prior work shows to be a powerful determinant of trust~\cite{ahn2024impact}.  

Future research should examine uncertainty visualization in longitudinal and real-world settings, where trust and reliance evolve over time~\cite{dietvorst2015algorithm,ahn2024impact}. Our findings suggest several concrete directions. First, providing richer model context, including training data, evaluation metrics, and explainable reasoning, may further enhance trust and decision quality. Second, adaptive strategies that layer binary and continuous cues could better accommodate differences in visualization literacy, while pairing uncertainty with outcome feedback and transparent model documentation may further calibrate trust. Third, studying the interaction between user expertise, cognitive load, and visualization complexity could clarify why inexperienced users sometimes prefer more elaborate visualizations. Finally, longitudinal studies in clinical workflows could assess how uncertainty visualizations affect long-term trust, reliance, and patient outcomes. Integrating these insights with prior work on multi-dimensional trust~\cite{ashoori2019ai} and AI transparency~\cite{lai2023towards} would help inform the design of interpretable, trustworthy AI systems that support safe, human--centered,
explainable and responsible AI systems in healthcare.

These findings highlight the multi-dimensional nature of trust in AI predictions and the importance of tailoring uncertainty visualization to both task context and user expertise. While our analyses reveal trade-offs between binary and continuous encodings and show that uncertainty cues alone are insufficient without contextual model information, they also point toward actionable strategies for design. In the following section, we distill these insights into a set of concrete design guidelines (Table~\ref{tab:guidelines}) to support the development of clinical decision support systems that foster calibrated reliance on AI predictions.  

\subsection{Design Guidelines for Visualizing AI Uncertainty}

\textcolor{blue}{Building on prior work in uncertainty visualization~\cite{kay2016ish,hullman2019al} and human--AI trust~\cite{ashoori2019ai,lee2004trust}, and grounded in our empirical findings from Studies~1 and~2, we propose six guidelines for designing uncertainty representations in prognostic AI systems. To clarify what these guidelines newly contribute, we explicitly distinguish (a) what prior literature already suggests, (b) what our study adds in the context of long-term AD prognosis, and (c) the implications for clinical AI design.}
\textcolor{blue}{ \paragraph{G1. Use continuous uncertainty to support reliability judgments.}
\textit{Prior work:} Uncertainty displays can improve calibrated reliance but may increase cognitive load~\cite{kay2016ish,hullman2019al}.  
\textit{Our findings:} Continuous uncertainty increased perceived reliability in Study~1 and increased all trust facets for experts in Study~2, despite lowering momentary trust for generalists.  
\textit{Implication:} Use continuous uncertainty when the goal is to support reflective assessment of model reliability, especially in prognostic contexts.}
\textcolor{blue}{\paragraph{G2. Use binary encodings cautiously to avoid overconfidence.}
\textit{Prior work:} Simplified encodings can support decisiveness but may mask uncertainty~\cite{poursabzisangdeh2021manipulating}.  
\textit{Our findings:} Binary uncertainty increased confidence for both generalists and experts, without corresponding gains in trustworthiness or reliability.  
\textit{Implication:} Reserve binary displays for rapid triage and pair them with reminders of residual uncertainty.}
\textcolor{blue}{\paragraph{G3. Combine uncertainty visualization with concise, layered model information.}
\textit{Prior work:} Lightweight model documentation helps users contextualize predictions~\cite{mitchell2019model}.  
\textit{Our findings:} Moderate model information increased confidence (Study~1), and experts requested model details even when continuous uncertainty was shown (Study~2).  
\textit{Implication:} Pair uncertainty visuals with brief model summaries and provide deeper documentation on demand.}
\textcolor{blue}{\paragraph{G4. Design and evaluate for multi-dimensional trust.}
\textit{Prior work:} Trust includes facets such as reliability, competence, and predictability~\cite{lee2004trust,ashoori2019ai}.  
\textit{Our findings:} Continuous uncertainty increased reliability but decreased momentary trust in Study~1. Nevertheless, in Study~2 it increased all trust facets while binary encodings inflated confidence.  
\textit{Implication:} Measure and design for separate trust facets rather than relying on a single global trust score.}
\textcolor{blue}{\paragraph{G5. Surface uncertainty prominently in long-term prognosis.}
\textit{Prior work:} Prognosis involves greater uncertainty and fewer feedback opportunities than diagnosis~\cite{jack2018nia}.  
\textit{Our findings:} Experts changed decisions in one-third of trials when uncertainty format changed, showing that uncertainty affects reasoning without destabilizing decisions.  
\textit{Implication:} In long-term predictions (e.g., 24-month AD progression), uncertainty must be highly salient.}
\textcolor{blue}{\paragraph{G6. Embed uncertainty visualization within broader transparency and validation.}
\textit{Prior work:} Trust increases more from evidence of model performance than from explanations alone~\cite{vereschak2024trust,bansal2021does}.  
\textit{Our findings:} Experts emphasized that ``uncertainty is not a proof of validity'' and requested external validation and training details.  
\textit{Implication:} Pair uncertainty displays with validation evidence, known failure modes, and concise explanations.}

\bigskip

\begin{table}[h!]
\centering
\small
\begin{tabular}{p{0.23\linewidth} p{0.32\linewidth} p{0.36\linewidth}}
\toprule
\textbf{Guideline} & \textbf{Known From Prior Work} & \textbf{What Our Study Adds} \\
\midrule

G1: Continuous for reliability &
Uncertainty displays can support calibrated reliance &
Reliability ↑ but momentary trust ↓ for generalists; experts show highest trust with continuous \\
\addlinespace[4pt]

G2: Binary with caution &
Simpler displays increase decisiveness &
Binary ↑ confidence for all users; evidence of potential expert overconfidence \\
\addlinespace[4pt]

G3: Pair uncertainty with model context &
Users benefit from brief information about how predictions are generated &
Moderate model information ↑ confidence; experts still request training and validation details \\
\addlinespace[4pt]

G4: Multi-dimensional trust &
Trust involves distinct facets (e.g., reliability, competence) &
Different encodings shift facets differently; confidence and trust diverge \\
\addlinespace[4pt]

G5: Make uncertainty salient in prognosis &
Prognostic tasks typically involve higher uncertainty than diagnosis &
Uncertainty format leads experts to reconsider decisions in ~1/3 of cases \\
\addlinespace[4pt]

G6: Integrate uncertainty with broader transparency &
Explanations alone do not ensure trust &
Experts: ``uncertainty is not validity''; they require external validation and model evidence \\
\bottomrule
\end{tabular}
\caption{Point-form summary of six guidelines, distinguishing prior knowledge from contributions unique to this work.}
\label{tab:guidelines}
\end{table}

\section{Conclusion}

\textcolor{blue}{Uncertainty visualization is increasingly proposed as a way to make AI systems more transparent, yet we still know relatively little about how specific encodings affect trust in AI and decision-making in long-term, high-uncertainty medical prognosis. This paper addressed that gap in the context of AI-supported AD progression, a domain where predictions are intrinsically uncertain, outcomes are delayed, and miscalibrated trust can have serious clinical consequences.}

Across two studies, we examined how different encodings (e.g. binary and continuous uncertainty) shape trust, confidence, and decision stability for predicted CN/MCI/AD transitions. In Study~1, participants reported more appropriately calibrated in-task trust and confidence with binary encodings than continuous encodings. While continuous encodings selectively increased perceived reliability. In Study~2, domain experts showed substantially more appropriately calibrated trust across all facets when uncertainty was visualized continuously, even though binary outputs again produced higher confidence, raising concerns about expert overconfidence in categorical displays. Experts also changed their decisions in about one-third of cases when viewing the same patient with different uncertainty encodings and emphasized that “uncertainty is not a proof of validity,” grounding their trust in external validation and model evidence.

These findings make three contributions. First, they provide empirical evidence that the effects of uncertainty visualization in AD prognosis are not uniform: visualization format and user expertise jointly determine whether uncertainty improves perceived reliability, inflates confidence, or both. Second, they demonstrate that trust, reliability, and confidence in AI predictions can diverge in systematic ways, underscoring the need to treat trust as multi-dimensional and to evaluate behavioral reliance alongside self-reports. Third, they yield six design guidelines that distinguish what was suggested by prior work from what emerges specifically from our studies: using continuous encodings to support reliability judgments, employing binary formats cautiously, pairing uncertainty with concise model context, designing for multi-dimensional trust, surfacing uncertainty prominently in long-term prognosis, and embedding uncertainty within broader transparency and validation.

In summary, our results show that uncertainty visualization in clinical AI is not simply a matter of ``adding error bars'', but of carefully matching encodings to user expertise, task horizon, and the kinds of trust that are desirable in practice. By grounding design guidelines in empirical trade-offs observed in AI-supported AD prognosis, this work advances both uncertainty visualization research and the design of explainable, trustworthy decision-support systems for neurodegenerative disease. Future work should extend these findings in longitudinal and human-in-the-loop clinical studies, where uncertainty cues, outcome feedback, and model updates interact over time to shape trust, reliance, and ultimately patient care.

\section{Acknowledgments}

\bibliographystyle{ACM-Reference-Format}
\bibliography{bibliography_ad}
\end{document}